\documentclass[twocolumn]{aastex63}

\usepackage{amsmath,amsxtra,amssymb,latexsym, amscd,amsthm}

\received{xxx}
\revised{xxx}
\accepted{xxx}
\submitjournal{ApJL}

\shorttitle{Vortex with Coagulation}
\shortauthors{Li et al.}

\begin{document}

\title{Planet-induced Vortices with Dust Coagulation in Protoplanetary Disks}

\correspondingauthor{Ya-Ping Li}
\email{leeyp2009@gmail.com}

\author[0000-0002-7329-9344]{Ya-Ping Li}
\affil{Theoretical Division, Los Alamos National Laboratory, Los Alamos, NM 87545, USA}
\author[0000-0003-3556-6568]{Hui Li}
\affil{Theoretical Division, Los Alamos National Laboratory, Los Alamos, NM 87545, USA}
\author[0000-0002-4142-3080]{Shengtai Li}
\affil{Theoretical Division, Los Alamos National Laboratory, Los Alamos, NM 87545, USA}
\author[0000-0002-1899-8783]{Tilman Birnstiel}
\affil{University Observatory, Faculty of Physics, Ludwig-Maximilians-Universit{\"a}t M{\"u}nchen, Scheinerstr. 1, 81679 Munich, Germany}
\affil{Exzellenzcluster ORIGINS, Boltzmannstr. 2, D-85748 Garching, Germany}
\author[0000-0002-9128-0305]{Joanna Dr\c{a}\.{z}kowska}
\affil{University Observatory, Faculty of Physics, Ludwig-Maximilians-Universit{\"a}t M{\"u}nchen, Scheinerstr. 1, 81679 Munich, Germany}
\author[0000-0002-1589-1796]{Sebastian Stammler}
\affil{University Observatory, Faculty of Physics, Ludwig-Maximilians-Universit{\"a}t M{\"u}nchen, Scheinerstr. 1, 81679 Munich, Germany}

\begin{abstract}
In this work, we study how the dust coagulation/fragmentation will influence the evolution and observational appearances of vortices induced by a massive planet embedded in a low viscosity disk by performing global 2D  high-resolution hydrodynamical simulations. 
Within the vortex, due to its
higher gas surface density and steeper pressure gradients, dust coagulation,
fragmentation and drift (to the vortex center) are all
quite efficient, producing dust particles ranging from micron to $\sim 1.0\ {\rm cm}$,
as well as overall high dust-to-gas ratio (above unity). 
In addition, the dust size distribution is quite non-uniform inside the vortex, 
with the mass weighted average dust size at the vortex center ($\sim 4.0$ mm)
being a factor of $\sim10$ larger than other vortex regions. 
Both large ($\sim$ mm) and small (tens of micron) particles contribute
strongly to affect the gas motion within the vortex. As such,
we find that the inclusion of dust coagulation has a significant impact
on the vortex lifetime and the typical vortex lifetime is about 1000 orbits. 
After the initial gaseous vortex is destroyed, the dust spreads into a ring 
with a few remaining smaller gaseous vortices with a high dust concentration and a large
maximum size ($\sim$ mm). At late time, the synthetic dust continuum images for the 
coagulation case show as a ring inlaid with several hot spots at 1.33 mm band, 
while only distinct hot spots remain at 7.0 mm. 
\end{abstract}

\keywords{accretion, accretion disks --- protoplanetary disks --- planets and satellites: formation --- planet-disk interactions --- methods: numerical}

\section{Introduction}
Vortices in protoplanetary disks may play essential roles in the early stage of planet formation due to their effectiveness in trapping dust, which could be an ideal place to trigger planetesimal formation \citep{Barge1995,Birnstiel2013,Casassus2013,Meheut2013,Zhu2012,Zhu2014}.  These vortices appeared as lopsided horseshoes, or crescent asymmetric features have been observed by several (sub-)mm observations, e.g., IRS 48 \citep{vanderMarel2013}, LkH$\alpha$ 330 \citep{Isella2013}, HD 142527 \citep{Muto2015}, MWC 758 \citep{Isella2010}, AB Aur \citep{Fuente2017}, SR 21 \citep{Perez2014}, and SAO 206462 \citep{Perez2014}. 
They can be formed by the Rossby wave instability (RWI; \citealt{Lovelace1999,Li2000,Li2001, Ono2018}) at the edges of a gap opened by a massive planet embedded in a low viscosity disk \citep{Li2005,Huang2018}, at the edges of the accretionally inactive dead zones \citep{Miranda2017,Regaly2017}, by a binary companion \citep{Calcino2019}, or by the baroclinic instability \citep{Klahr2003,Raettig2013,Lyra2014}, .

The long term evolution of vortices  has been studied by \citet{Fu2014a}, which suggests that a low gas viscosity is required  to sustain vortices to thousands and up to $10^{4}$ orbits (see also \citealt{deValBorro2007,Hammer2017}). \citet{Fu2014b} further found that the feedback effect from the dust (the back reaction of dust onto gas) in 2D simulations can reduce the lifetime of the vortices by a factor of up to 10 for a variety of initial dust-to-gas ratios and single dust particle sizes.
It has also been found that vortices in 3D disks are subject to
``elliptic instability", which reduces the viability of these vortices as dust traps  \citep{Lithwick2009,Lesur2009,Barge2016,Lin2018}. 
However, the dust size growth during the whole evolution of the disk, which is implemented only recently in the global 2D disk model (\citealt{Drazkowska2019,Laune2020}), has not been considered for all these studies.
\citet{Drazkowska2019} studied the dust distribution in the vicinity of a Jupiter-mass planet embedded in a relatively high viscosity disk, while \citet{Laune2020} explored the ring morphology affected by coagulation in the parameter regime of a low viscosity and a low planet mass.

As the dust feedback becomes important when the dust-to-gas ratio approaches unity, which increases with time in the vortex region due to the radial drift and collecting process. 
Both of these processes become faster with a larger Stokes number (i.e., $\propto{\rm St^{-1}}$, St is defined in Equation~\ref{eq:st}) or a larger dust size in the small Stokes number regime, although the collecting process is much faster than the radial drift \citep{Surville2016}.
Therefore, the coagulation, which controls the dust size growth, can play an important role in controlling the efficiency of dust feedback. The fragmentation, however, distributes the surface density from large particles into small ones, thus the increase of the dust-to-gas ratio cannot be so efficient compared with single species model. The efficiency of dust feedback, and then the evolution of large-scale vortices, thus results from the complex interplay between the dust coagulation/fragmentation and radial drift.
Here we include the dust coagulation/fragmentation to study its effect on the evolution of planet-induced vortices and observational appearances in protoplanetary disks.

The rest part of the \textit{Letter} is organized as follows. Our coagulation and single species models are described in Section~\ref{sec:model}. We then present the results and discuss the observational implication in Section~\ref{sec:results}, and conclude in Section~\ref{sec:conclusions}.

\section{Methods}\label{sec:model}

Similar to \citet{Drazkowska2019} and \citet{Laune2020}, where they explored different regimes of planet mass or viscosity, we study the effect of coagulation on the vortex evolution induced by a massive planet embedded in a low viscosity protoplanetary disk.
As inferred from observations for IRS 48 \citep{vanderMarel2013}, a 5 $M_{\rm J}$ (where $M_{\rm J}$ is the Jupiter mass) planet is assumed to orbit a $2\ M_{\odot}$ (where $M_{\odot}$ is the solar mass) star on a fixed circular orbit at a radius of $20\ {\rm au}$. A low viscosity parameter $\alpha_{\rm vis}=7\times10^{-5}$ throughout the disk is adopted to sustain the gaseous vortex \citep{Fu2014a}. We assume a $\alpha$-prescription for the gas kinematic viscosity $\nu_{\rm g}=\alpha_{\rm vis} c_{\rm s}h_{\rm g}$ \citep{Shakura1973}.

We choose an exponential decay profile for the initial gas surface density $\Sigma_{\rm g}(r)$ as 
\begin{equation}\label{eq:gas}
  \Sigma_{\rm g}(r)=\Sigma_{0}\left(\frac{r}{r_{\rm c}}\right)^{-\gamma}\exp\left[-\left(\frac{r}{r_{\rm c}}\right)^{2-\gamma}\right],
\end{equation}
where $r_{\rm c}=80~\rm au$ and $\gamma=0.8$. The normalization of gas surface density $\Sigma_{0}=1.3\ {\rm g\ cm^{-2}}$. The disk extends from 8 au to 320 au. Disk self-gravity in gas and dust is not included due to a low disk mass of $M_{\rm disk}=4.5\times10^{-3}\ {M_{\odot}}$, or equivalently a large minimum Toomre $Q$ parameter across the disk $\sim150$ initially, as suggested by previous works \citep{Lovelace2013,Zhu2016}.
The locally isothermal sound speed $c_{\rm s}$ is chosen as $\frac{c_{\rm s}}{v_{\rm K}}=\frac{h_{\rm g}}{r}=h_{0}\left(\frac{r}{r_{0}}\right)^{0.25}$,
where $v_{\rm K}(r)$ is the local Keplerian speed, $r_{0}=20$ au, and 
$h_{0}=0.06$. 
This corresponds to a disk temperature profile as $T=89.0(r/r_{0})^{-0.5}{\ \rm K}$. We adopt an isothermal equation of state $P=c_{\rm s}^{2}\Sigma_{\rm g}$ for the gas component, 
where $P$ is the vertically integrated gas pressure.
The gas and dust fluids for the coagulation model are evolved following the conservation of mass, radial, 
and angular momentum equations  \citep{Li2019b}.

The dust feedback, i.e., drag forces between the gas and dust, are incorporated into the momentum equation for both the gas and dust \citep{Fu2014b}. The drag force $f_{\rm d}^{i}$ for a dust species with its size $a^{i}$ is defined as
\begin{equation}\label{eq:drag}
\mathbf{f}_{\rm d}^{i}=\frac{\Omega_{\rm K}}{{\rm St}^{i}}(\mathbf{v}_{\rm g}-\mathbf{v}_{\rm d}^{i}),
\end{equation}
where $\Omega_{\rm K}$ is the Keplerian angular velocity.  ${\rm St^{i}}$ and $\mathbf{v}_{\rm d}^{i}$ are the 
Stokes number,  dust velocity for species $i$, respectively. $\mathbf{v}_{\rm g}$ is the gas fluid velocity.  
We have included Epstein and Stokes regimes for the aerodynamic drag between gas and dust. In the Epstein regime for most region of the disk, the Stokes number of the particle with a dust radius $a$ in the mid-plane of the disk is defined as 
\begin{equation}\label{eq:st}
  {\rm St} = \frac{\pi\rho_{\rm s}a}{2\Sigma_{\rm g}},
\end{equation}
where $\rho_{\rm s}=0.8\ {\rm g\ cm^{-3}}$ is the internal density of the dust particles. For our disk parameters, ${\rm St}(r_{0})=0.16$ with a dust size of 4.0 {\rm mm}. The dust size corresponding to a unity Stokes number is $ a_{\rm St=1}=\frac{2\Sigma_{\rm g}}{\pi\rho_{\rm s}}$, which is $2.4\ {\rm cm}$ at $20\ {\rm au}$ for our initial gas profile.

To understand how the feedback force influences the dust evolution,  we adopt the approach of \citet{Takeuchi2002} to examine its dependence on ${\rm St}$ before the back reaction becomes important for the gas dynamics. 
The radial velocity of the dust is
\begin{equation}\label{eq:vr_dust}
v_{{\rm d},r}=\frac{v_{{\rm g},r}+2{\rm St}\Delta v_{{\rm g},\phi}}{1+{\rm St^2}},
\end{equation}
where $\Delta v_{{\rm g},\phi}=v_{{\rm g},\phi}-v_{\rm K}\simeq-1/2\eta v_{\rm K}$, the second equality applies when $\eta\ll1$, 
where $\eta=-\frac{c_{\rm s}^2}{v_{\rm K}^2}\frac{d\log P}{d\log r}$,
and
\begin{equation}\label{eq:vr_gas}
v_{{\rm g},r}=-\frac{3}{\Sigma_{\rm g}\sqrt{r}}\frac{\partial}{\partial r}(\Sigma_{\rm g }\nu_{\rm g}\sqrt{r})
\end{equation}
is the radial velocity of the gas when there is no dust back reaction, which is on the order of $\sim\alpha_{\rm vis}(c_{\rm s}/v_{\rm K})^{2}v_{\rm K}$. When ${\rm St}\ll1$, the radial velocity of dust is reduced to $v_{{\rm d},r}\simeq v_{{\rm g},r}+2\Delta v_{{\rm g},\phi}{\rm St}$.
The radial force is then expressed as:
\begin{equation}\label{eq:fdr}
f_{\mathrm{d},r}= \frac{\mathrm{St}v_{{\rm g},r}-2\Delta v_{{\rm g},\phi}}{\mathrm{St}^{2}+1}\Omega_{\rm K}.
\end{equation}
When $\rm St \ll1$, $f_{\mathrm{d},r}\simeq-2\Delta v_{{\rm g},\phi}\Omega_{\rm K}$ is independent on $\rm St$.  We find that $f_{\mathrm{d},r}$ usually dominates over $f_{\mathrm{d},\phi}$ if ${\rm St}\lesssim1$. The effective drag force exerted on gas from each species is then $\mathbf{F}_{\rm d}^{i}=\mathbf{f}_{\rm d}^{i}\Sigma_{\rm d}^{i}/\Sigma_{\rm g}$ for each species, where $\Sigma_{\rm d}^{i}$ is the dust surface density \citep{Fu2014b,Li2019b}. The total drag force $\mathbf{F}_{{\rm d}}$ for all dust species $\mathbf{F}_{\rm d}^{i}$ is the summation of $\mathbf{F}_{\rm d}^{i}$ over $i$ if multiple dust species is included. This value can be significant if the pressure gradient parameter $\eta$ in the vortex region becomes large, and $\mathbf{F}_{{\rm d}}$ can be comparable to pressure forces $(1/\Sigma_{\rm g})dP/dr$ if $\Sigma_{\rm d}/\Sigma_{\rm g}\sim1$.

We run four models to quantify the effect of coagulation on the vortex evolution. One includes 2D coagulation, and other two are for a single dust species. For the coagulation run, the details have been described in \citet{Li2019b,Drazkowska2019} and \citet{Laune2020}. 
Only 1.0 $\mu$m sized dust particles are included in the disk initially.  The dust size distribution is resolved with 151 dust species covering sizes between $1.0\ {\rm \mu m}$ and $100\ {\rm cm}$. 
Collisional outcomes include sticking (fragmentation) when impact speeds for collisions are below (above) a critical speed of $v_{\rm f}=10\ {\rm m\ s^{-1}}$. 
Due to the computational expense of dust coagulation, we implement a sub-stepping routine and call the coagulation solver every 50 hydro time steps\footnote{Due to the existence of asymmetric features, we have tested the coagulation model with 25 hydro time steps using a low resolution ($1024\times1024$), and find that it can slightly speed up the destruction of the large-scale vortex. But note that in our high resolution run, the hydro step is much smaller and 50 hydro steps correspond to 0.02 orbit, which can resolve the fine structures within the vortex. We, therefore, expect this sub-stepping does not change our results significantly.}. 
We use a turbulence parameter $\alpha_{\rm t}=10^{-3}$  for the dust coagulation/fragmentation, which is different from the gas viscosity $\alpha_{\rm vis}$ \citep{Carrera2017}. In most cases, we should expect that $\alpha_{\rm t} < \alpha_{\rm vis}$, where the turbulence that stirs dust can also induce gas accretion.
One main reason for adopting $\alpha_{\rm t} > \alpha_{\rm vis}$ here is that we need a low $\alpha_{\rm vis}$ to trigger RWI for vortex formation, while a larger $\alpha_{\rm t}$ can avoid an extremely large dust size due to coagulation.
Another possibility is that the mid-plane $\alpha_{\rm t}$ could be determined by the local instability (e.g., vertical shear instability), while $\alpha_{\rm vis}$ controls the \textit{global} viscosity for the disk accretion, which can be determined by other MHD processes (e.g., \citet{Bai2013}). The dust turbulence parameter $\alpha_{\rm t}=10^{-3}$ we adopt is close to the inferred value from observations \citep{Flaherty2017}. The region with a smaller $\alpha_{\rm vis}$ could correspond to the viscously inactive zone, where the vortex tends to be formed, while the disk global viscous evolution for the region far from the vortex, even with a larger $\alpha_{\rm vis}\sim10^{-2}$, should be unaffected within a timescale of $\sim1000$ orbits.

For the run with a single dust species, we have two runs with the dust size fixed at $a=4.0\ {\rm mm}$ or $a=0.2\ {\rm mm}$. The size $a=4.0\ {\rm mm}$ is close to the $\Sigma_{\rm d}-$weighted dust size during the evolution of our coagulation model, while $a=0.2\ {\rm mm}$ is the commonly used dust size for the single species run.
For all cases, the surface density distribution of dust
follows the radial profile of the gas with an initial radial-independent dust-to-gas mass ratio of $0.01$ at the initial stage.
To mimic the coagulation run, we also have another run with 5 species of dust logarithmically uniform spaced between $1.0\ \mu{\rm m}$ and $4.0\ {\rm mm}$ with an initial MRN distribution \citep{Mathis1977} to examine the long-term evolution behaviour. Note that dust coagulation/fragmentation is not considered for this model.

We solve the 2D hydrodynamics equations with \texttt{LA-COMPASS} \citep{Li2005,Li2009,Fu2014b,Li2019b} in a logarithmically radial grid of $n_{r}=4096$, and a uniform azimuthal grid of $n_{\phi}=3456$. With a such high resolution to capture the vortex structures \citep{Fu2014b}, the coagulation run is quite computational expensive. Specifically, it takes about 1 million CPU hours for our coagulation model, which makes a parameter study unrealistic. 
We keep the gas density constant at the inner and outer boundary. An outflow boundary condition are imposed on the dust inner/outer boundary (\citealt{Drazkowska2019,Li2019a}). 

\section{Results}\label{sec:results}

\subsection{Coagulation Model}

\begin{figure}[htbp]
\begin{center}
\includegraphics[width=0.5\textwidth]{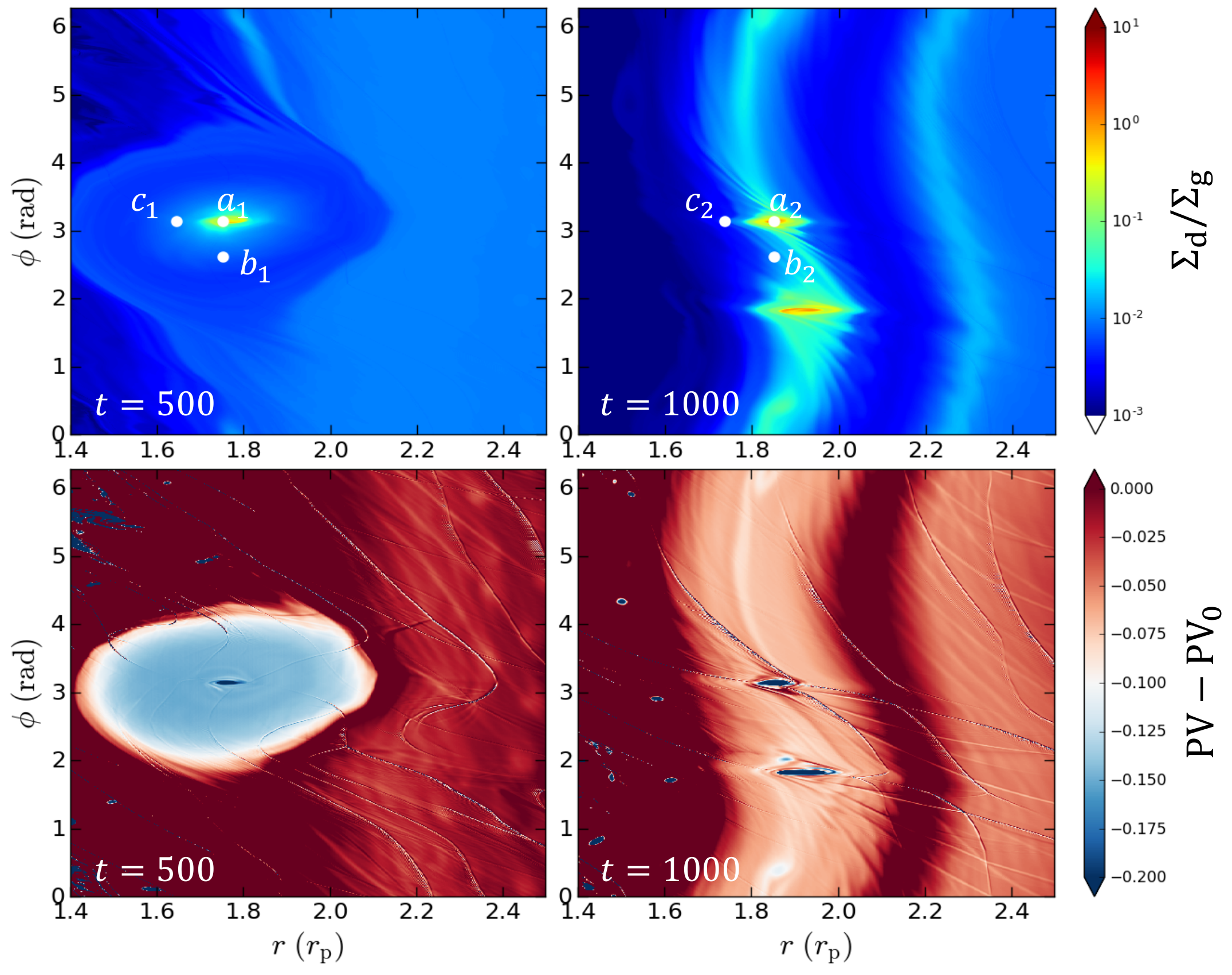}
\end{center}
\caption{Gas and dust dynamics around the gaseous vortex region for the coagulation run. Upper: total $\Sigma_{\rm d}/\Sigma_{\rm g}$ for all dust species at $500$ (left) and $1000$ (right) orbits. Lower: PV subtracted from the initial value at $500$ (left) and $1000$ (right) orbits. }\label{fig:coag}
\end{figure}

We first show the gas and dust dynamics of coagulation model.
The massive planet can quickly carve out a clean gap around the planet location. The outer edge of the gap ($r\sim1.8$) becomes Rossby wave unstable. The multiple vortices produced at the early stage quickly merge into a large-scale gaseous vortex, which can be seen from the snapshot for gas potential vorticity (${\rm PV}=(\bigtriangledown\times \mathbf{v}_{\rm g})/\Sigma_{\rm g}$) at $t=500$ orbits. The PV contours subtracted from its initial value are shown in the lower panels of Figure~\ref{fig:coag}. At $t=500$, PV is still relatively smooth with a minimum at the vortex center. 

\begin{figure}[htbp]
\begin{center}
\includegraphics[width=0.4\textwidth]{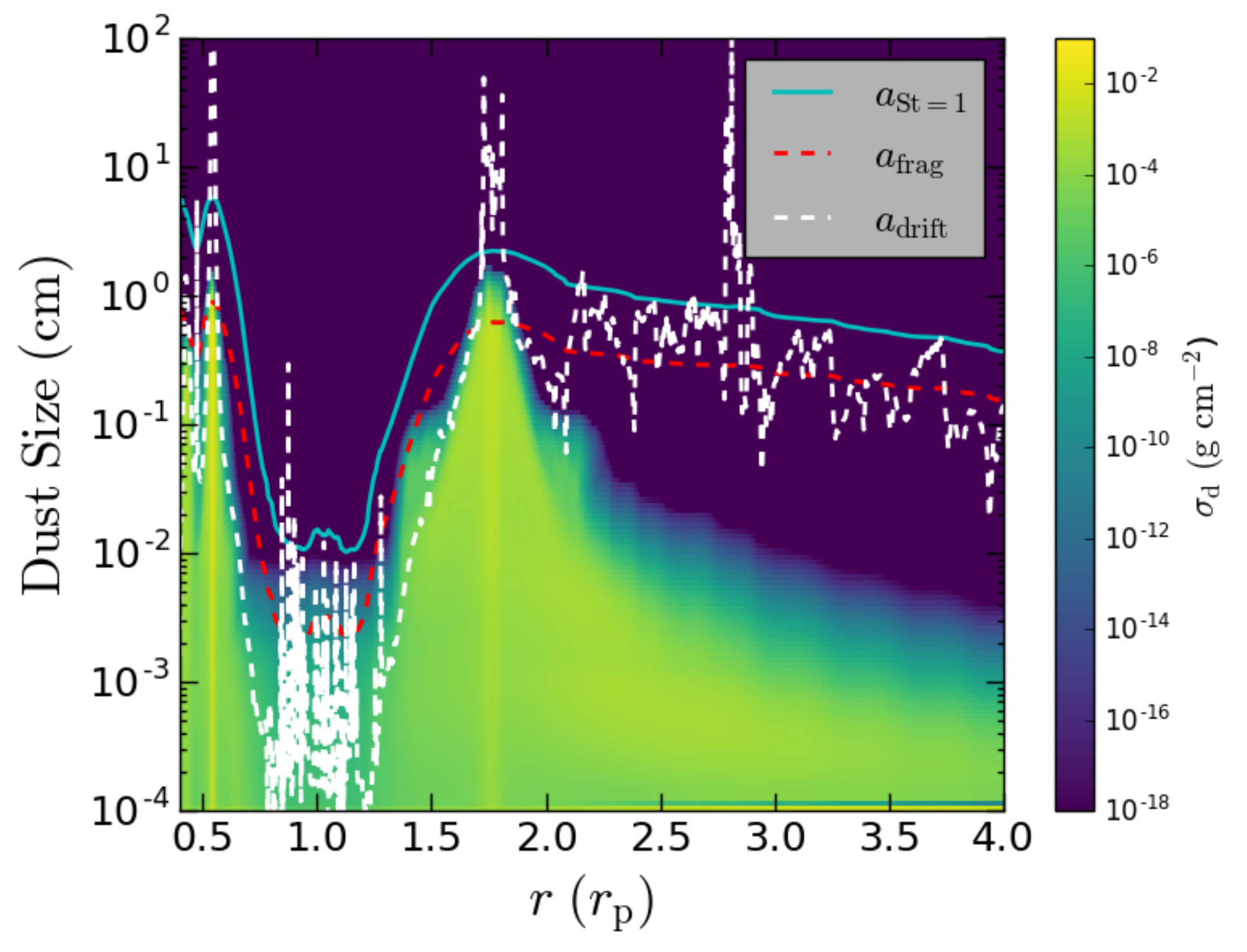}
\includegraphics[width=0.4\textwidth]{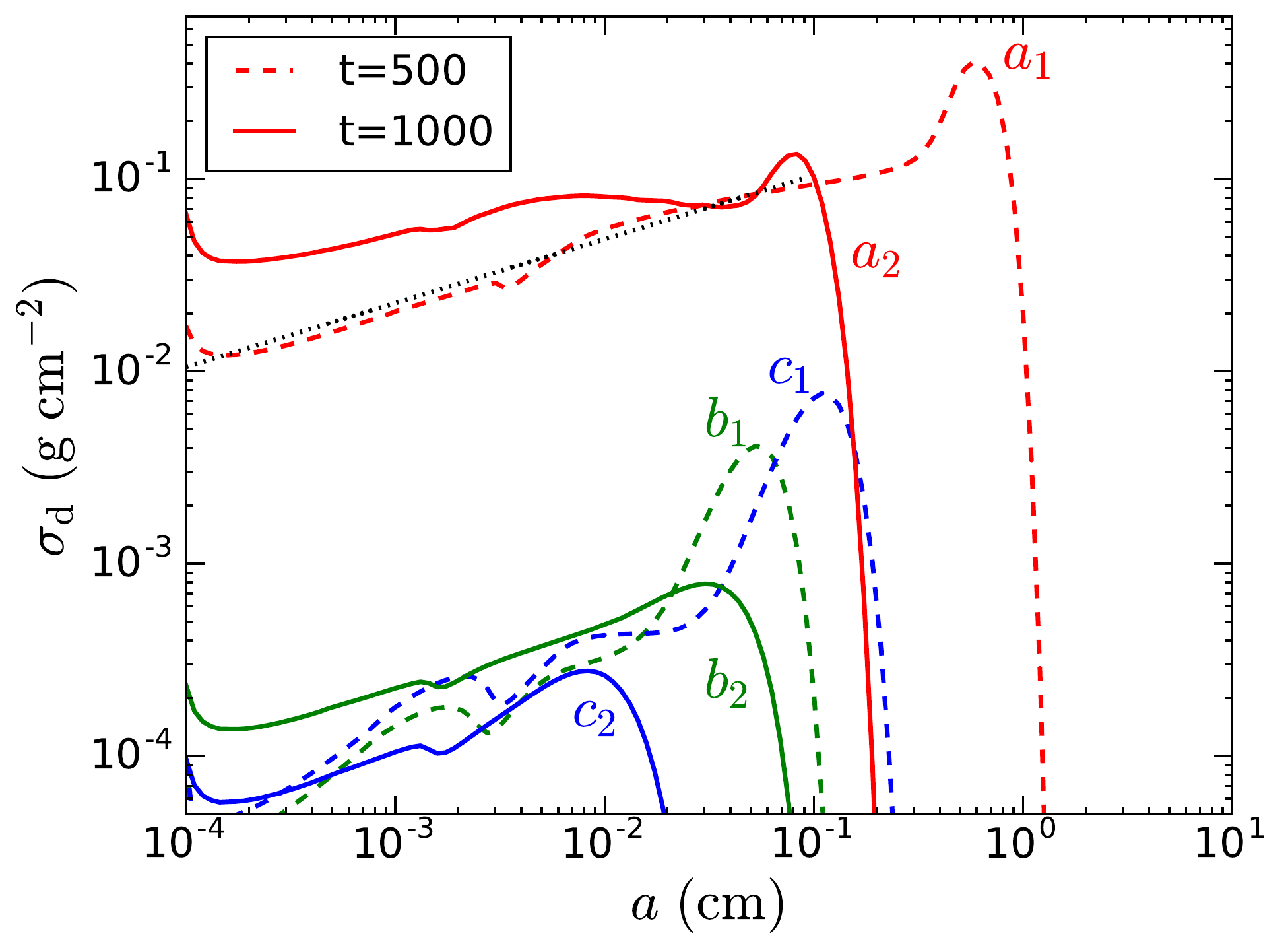}
\includegraphics[width=0.4\textwidth]{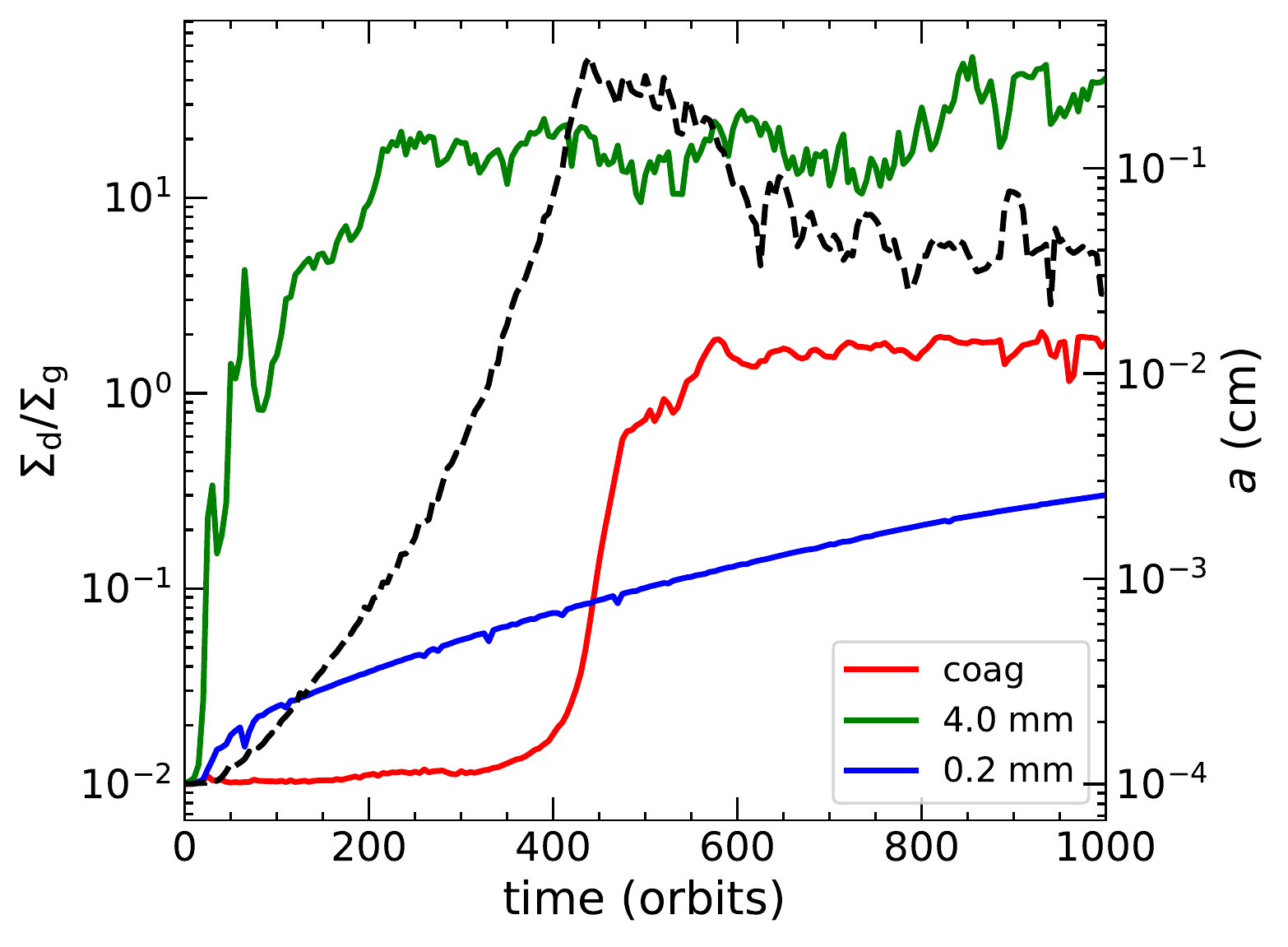}
\end{center}
\caption{Upper: the azimuthal-averaged dust surface density as a function of dust size and radial distance from the star for the time of 500 orbits. Different lines correspond to different size limits. The oscillation of $a_{\rm drift}$ is related to the azimuthal-averaging of the wobbling disk.  Middle: dust surface density as a function of dust size at three locations, which are marked in the upper panels of Figure~\ref{fig:coag}. The black dotted line is a power-law fit to the dashed red line with a slope of 0.33. The red lines correspond to the location of the maximum dust surface density at each time. Bottom: the time evolution of the dust-to-gas ratio $\Sigma_{\rm d}/\Sigma_{\rm g}$ at the vortex center (i.e., the location of maximum $\Sigma_{\rm d}$) for three models (color solid lines). The black dashed line shows the $\Sigma_{\rm d}$-weighted dust size $a$ evolution at the vortex center for the coagulation model. }\label{fig:coag_dis}
\end{figure}

As the particles drift from the outer region of disk, they will be collected into the vortex region, which is associated with the gas bump. The particles grow in size in accompany with the drift.  Due to the small particle size in the outer region of disk where particles drift inward (i.e., less than $0.2\ {\rm mm}$, see the upper panel of Figure~\ref{fig:coag_dis}), only a small fraction of the total dust mass ($\sim17\%$, or $2.0\ M_{\oplus}$, where $M_{\oplus}$ is the Earth mass) can be collected into the vortex region (e.g., the radial band of $r\sim[1.6,2.0]$). 
The azimuthal-averaged dust size distribution at 500 orbits are shown in the upper panel of Figure~\ref{fig:coag_dis}. 
We can see that the dust growth is mainly limited by the fragmentation $a_{\rm frag}$ \citep{Birnstiel2012,Li2019b} and radial drift barrier $a_{\rm drift}$ \citep{Birnstiel2012} calculated using the azimuthal gas surface density profile. At the outer edge of the gap, the maximum dust size can only be a few tens of $\mu{\rm m}$.

The dust mass can then spiral inward into the vortex center.
The dust fragmentation and drift barrier in the center is also larger due to the existence of the gas bump, leading to an increase of the maximum dust size to $\sim1.0\ {\rm cm}$ while sinking into the center (point $a_{1}$, corresponding to the location of the maximum $\Sigma_{\rm d}$) from the edge (points $b_{1}$ and $c_{1}$) of the vortex, as shown in middle panel of Figure~\ref{fig:coag_dis}. The size distribution is quite non-uniform inside the vortex, with the $\Sigma_{\rm d}-$weighted dust size at the vortex center ($\sim4.0\ {\rm mm}$) being a factor of $\sim10$ larger than other vortex regions.
The size growth can also speed up the collecting process of the particles due to its dependence on Stokes number \citep{Surville2016}. Therefore, both $\Sigma_{\rm d}$ and $\Sigma_{\rm d}/\Sigma_{\rm g}$ increase rapidly toward the center region. The dust surface density of small particles around the vortex region is also enhanced accompanied with the big ones, as shown by the dashed line in the middle panel of Figure~\ref{fig:coag_dis},  because they are created by fragmenting the big ones.
A distinct maximum for $\Sigma_{\rm d}/\Sigma_{\rm g}$ (or dust surface density) is then formed at the center as shown in the upper left panel of Figure~\ref{fig:coag} (see also \citealt{Fu2014b,Crnkovic2015}). The scale of the small dust clumps shown in the upper panels of Figure~\ref{fig:coag} is mainly determined by the dust diffusivity \citep{Chang2010}.

The time evolution of  $\Sigma_{\rm d}/\Sigma_{\rm g}$ is shown in the lower panel of Figure~\ref{fig:coag_dis}, which clearly demonstrates two stages of  evolution. The dust grows from the initial size of $1\ \mu{\rm m}$ to the maximum value by the time of 400 orbits, which is a factor of two longer than the estimate by \citet{Birnstiel2012} and \citet{Laune2020}. This discrepancy may be due to the fact that the relative velocity of the dust is dominated by radial drift rather than turbulence in the outer disk. In this first stage, the increase of $\Sigma_{\rm d}/\Sigma_{\rm g}$ is quite inefficient due to the small Stokes number. The efficient dust collecting process starts after the dust grows to the maximum size.

As particles collecting process proceeds, the total dust-to-gas ratio $\Sigma_{\rm d}/\Sigma_{\rm g}$ is higher than unity in the center of the vortex at 600 orbits and gets saturated afterwards, as shown in the lower panel of Figure~\ref{fig:coag_dis}. 
This collecting timescale ($\sim200$ orbits) is roughly consistent with the analytical estimate of $\sim170$ orbits by \citet{Surville2016} ($2\tau_{1/2}$ defined in Equation~(38) therein) with an initial Stokes number of 0.01. It is thus the combination of dust size growth and collecting processes that determine that the timescale for the increase of $\Sigma_{\rm d}/\Sigma_{\rm g}$ to unity, which finally controls the lifetime of the gaseous vortex.
The feedback of dust onto gas can trigger the vortex streaming instability \citep{Surville2016}, which produces fluffy ``fingers" features around the center of the vortex as discovered in \citet{Fu2014b} and \citet{Crnkovic2015}. These fluffy features can result in the elongation in the azimuthal direction, and finally destroys the large-scale vortex due to the ``heavy core" instability \citep{Chang2010}, which disperses the large-scale vortex into the whole azimuthal domain as seen from the right panels of Figure~\ref{fig:coag}. There exists some substructures in the gas surface density as seen from the PV plot, which can still trap the dust into two bumpy regions. The dust is also stretched into an elongated ring as the PV pattern. 
The maximum dust size in the PV minimum decreases to a smaller value ($\sim 1.0\ {\rm mm}$), as shown in the middle and lower panels of Figure~\ref{fig:coag_dis}, because the gas bump becomes much shallower.
A dusty ring is formed after the destruction of the vortex, with the wobbling features related to the very massive planet.

\subsection{Comparison with Single Species Models}
\begin{figure}[htbp]
\begin{center}
\includegraphics[width=0.5\textwidth]{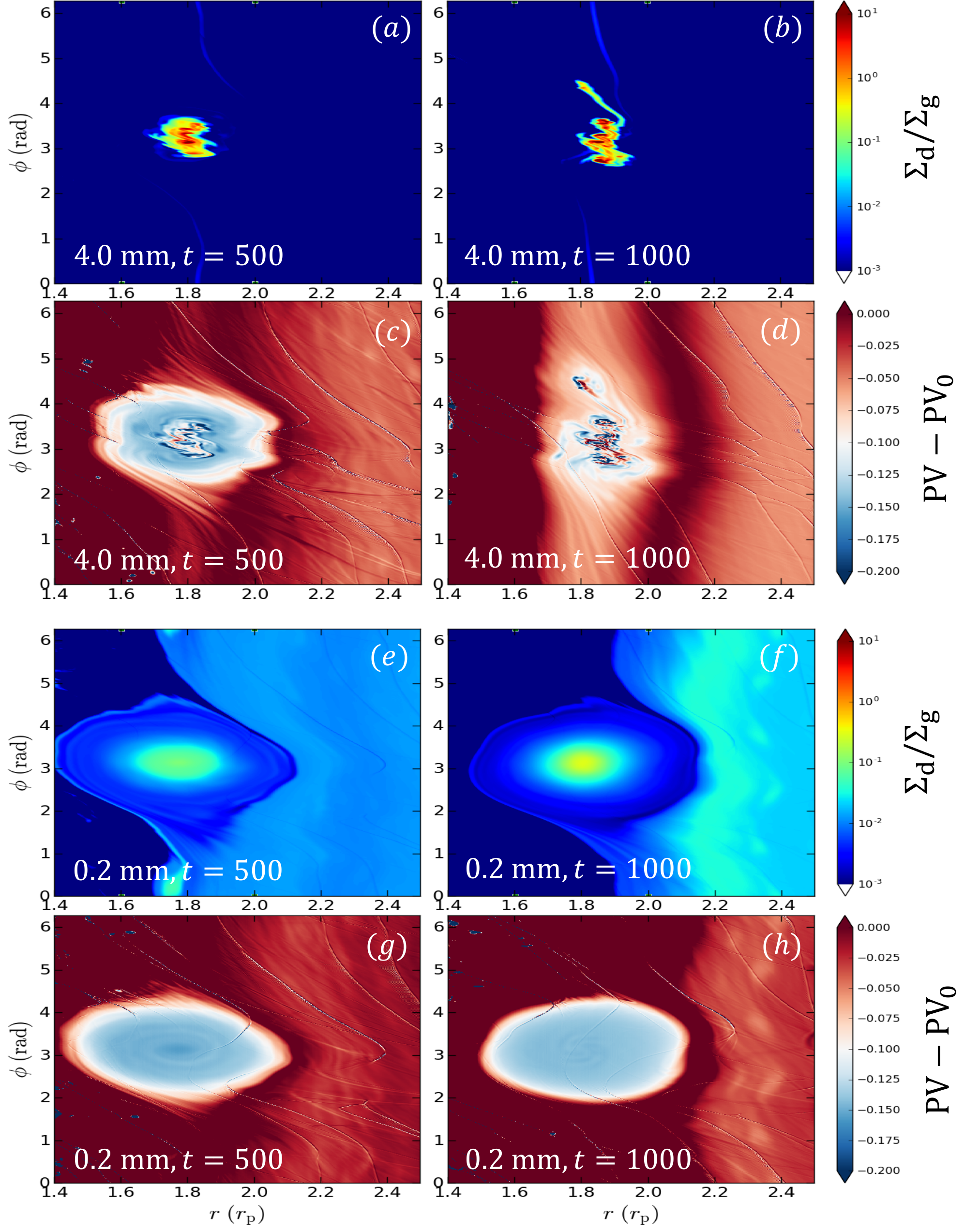}
\end{center}
\caption{$\Sigma_{\rm d}/\Sigma_{\rm g}$ and $\rm PV$ around the gaseous vortex region for the single species runs with dust sizes of $a=4.0\ {\rm mm}$ (panels $a-d$) and $a=0.2\ {\rm mm}$ (panels $e-h$) at the 500 (left panels) and 1000 (right panels) orbits. }\label{fig:sig_sdust}
\end{figure}

The effect of dust feedback from the single species on the evolution of the gaseous vortex has been studied by several authors \citep{Fu2014b,Crnkovic2015,Surville2016}. They have shown that a large dust size (equivalently, a large initial Stokes number) can shorten the lifetime of the vortex. Here we mainly use these single species runs as a comparison to quantify the effect of dust coagulation on the evolution of vortex. We keep all other model parameters being the same except adopting a single dust size of $a=4.0\ {\rm mm}$ and $a=0.2\ {\rm mm}$ for two runs.
For the case of dust size $a=4.0\ {\rm mm}$,  the large-scale vortex 
has already disappeared at $\sim 500$ orbits, as shown in the panel (c) of Figure~\ref{fig:sig_sdust}. The PV contour starts to become fluffy at the very  early stage (e.g., before $t=500$ orbits), and is quite turbulent at 1000 orbits which leads to the elongation in the whole azimuthal direction. The $\Sigma_{\rm d}/\Sigma_{\rm g}$ contour is quite clumpy (panels a--b), which is related to the gas substructures in the same region, as can be indicated from the PV plot shown in panels (c--d) of Figure~\ref{fig:sig_sdust}. 

When the dust size decreases to $a=0.2\ {\rm mm}$, the gas vortex can survive for a much longer time, as shown in the panels (e--h) of Figure~\ref{fig:sig_sdust}. The gaseous vortex is still quite strong, and $\Sigma_{\rm d}/\Sigma_{\rm g}$ contour is much smoother up to 1000 orbits.

The remarkable difference of the lifetime between two single species models can also be understood from the time evolution of the dust-to-gas ratio at the vortex center shown in the lower panel of Figure~\ref{fig:coag_dis}. The collecting process for the $a=4.0\ {\rm mm}$ model is quite efficient compared with the coagulation run due to the absence of the initial size growth process, and also much faster than the $a=0.2\ {\rm mm}$ model. The later is simply due to the large difference of dust Stokes number for two models ($\propto{\rm St}^{-1}$).

\subsection{Dust Feedback Efficiency}

\begin{figure}[htbp]
\begin{center}
\includegraphics[width=0.4\textwidth]{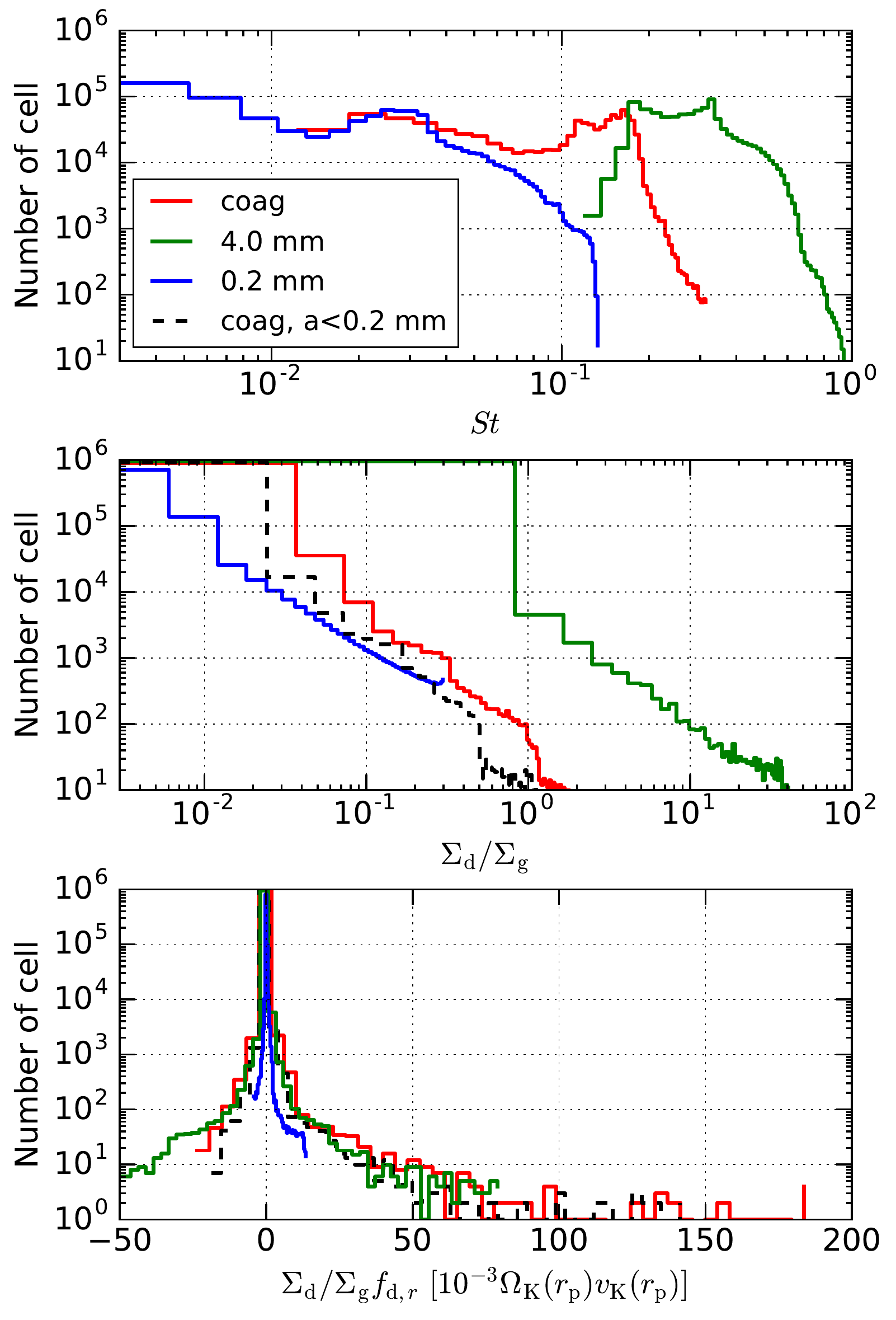}
\end{center}
\caption{Histogram of different quantities with uniform binning for three runs at the region between $r\sim[1.6, 2.0]$ where the gaseous vortex is located. All these quantities are measured at $1000$ orbits. Upper: the distribution of Stokes number. For the coagulation run, we use the maximum dust size, which is close to the turning point in the dust size distribution to calculate the Stokes number. Middle: $\Sigma_{\rm d}/\Sigma_{\rm g}$ distribution. Lower: radial feedback force $F_{{\rm d},r}=\Sigma_{\rm d}/\Sigma_{\rm g}f_{{\rm d},r}$ distribution. For the coagulation run, $F_{{\rm d},r}$ is the summation of $F_{{\rm d},r}^{i}$ over all dust species $i$.  The dashed lines in the middle and lower panels show $\Sigma_{\rm d}/\Sigma_{\rm g}$ and forces contributed by small dust particles ($\lesssim0.2\ {\rm mm}$). We can see that small particles contribute a large fraction to the total forces. }\label{fig:hist}
\end{figure}

We have found that the vortex lifetime in the coagulation model is in between the small and large dust runs. Since it is the efficient dust feedback process that triggers the vortex streaming instability \citep{Surville2016}, which indicates the starting point of vortex destruction by the ``heavy core" instability \citep{Chang2010},
we plot the statistical properties in the radial band of $r\sim[1.6, 2.0]$, where the gaseous vortex is located, for several physical quantities in Figure~\ref{fig:hist}, to demonstrate different feedback efficiencies in the vortex region.

When the dust coagulation is included, the distribution of Stokes number calculated using the maximum dust size\footnote{The maximum dust size is approximated by the turning point in the size distribution as shown in the middle panel of Figure~\ref{fig:coag_dis}.} in each cell falls in between the Stokes numbers corresponding to the two single species runs.  
Although most regions still have a small dust size ($\lesssim0.2\ {\rm mm}$) and a low Stokes number ($\lesssim0.1$) as shown in the middle panel of Figure~\ref{fig:coag_dis},
the total $\Sigma_{\rm d}/\Sigma_{\rm g}$, summed over all dust species, can be close to unity, which is much higher than that of $a=0.2\ {\rm mm}$ run, even though the dust mass accumulated in the vortex region is less than that of $a=0.2\ {\rm mm}$ run. This is because the vortex region, especially at its center, has a larger particle size as discussed above, which leads to an efficient collecting process into a small region within the vortex.
Therefore, it is radial and azimuthal drift that delivers large particles to the vortex center, but it is the fragmentation of those particles that subsequently boosts the coupling between dust and gas by increasing the dust surface area.

We further show how this can affect the total feedback force in the vortex region.
$f_{{\rm d},r}^{i}$ becomes a large constant of $\eta v_{\rm K}\Omega_{\rm K}$ for a small Stokes number as shown in Equation~(\ref{eq:fdr}), which is comparable to pressure forces if $\Sigma_{\rm d}/\Sigma_{\rm g}\sim1$.
Such a large $f_{{\rm d},r}$, together with the fact that $\Sigma_{\rm d}/\Sigma_{\rm g}$ is close to unity, can contribute to strong feedback forces $\Sigma_{\rm d}/\Sigma_{\rm g}f_{{\rm d},r}$, as shown in the lower panel of Figure~\ref{fig:hist}. We  further find that the feedback forces from small particles with $a\lesssim0.2\ {\rm mm}$ contribute comparably to total forces ($\sim70\%$ of total forces when summed over the central tiny PV minimum region) as indicated by the dashed lines, which is due to the comparable  contribution of $\Sigma_{\rm d}/\Sigma_{\rm g}$ from these small particles. 
Therefore, it is the high $\Sigma_{\rm d}/\Sigma_{\rm g}\sim1$ that directly initiates the strong feedback effect and then destroys the vortex, while the large Stokes number plays the role in facilitating the increase of $\Sigma_{\rm d}/\Sigma_{\rm g}\sim1$.

The models with only small dust as in our single species run with $a=0.2\ {\rm mm}$, even though they can have a large drag force $f_{\rm r}$ for one dust species,  cannot enhance $\Sigma_{\rm d}/\Sigma_{\rm g}$ efficiently in the vortex region, and therefore results in an inefficient feedback process, shown as blue lines in Figure~\ref{fig:hist}.
When the dust size for the single species run becomes much larger (i.e., $a=4.0\ {\rm mm}$),  $\Sigma_{\rm d}/\Sigma_{\rm g}$ increases significantly in the vortex region due to the efficient collecting process of dust, which can compensate the decrease of $f_{{\rm d},r}$ with Stokes number, and finally leads to the strong feedback force shown in Figure~\ref{fig:hist}.

\begin{figure*}[htbp]
\begin{center}
\includegraphics[width=0.75\textwidth]{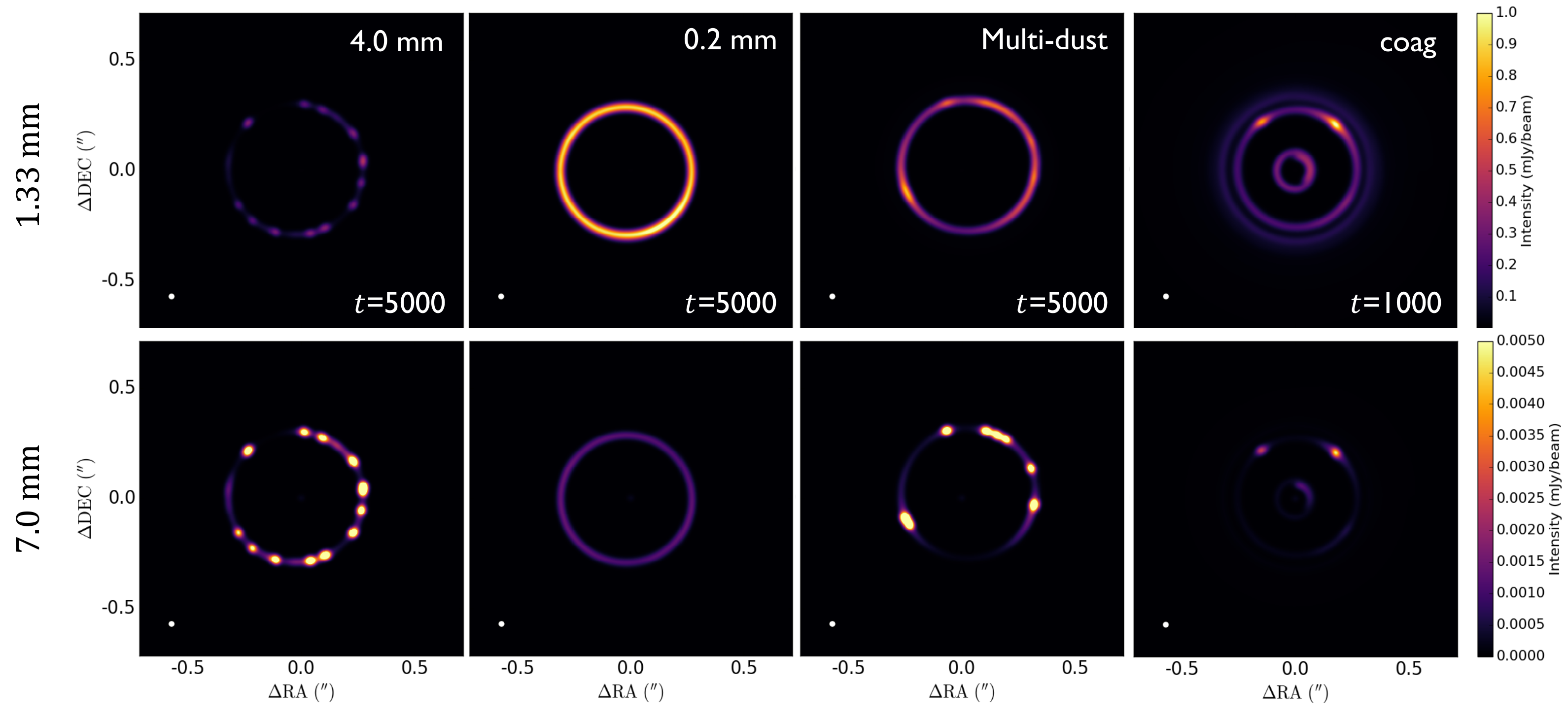}
\end{center}
\caption{1.33 mm (upper panels) and 7.0 mm (lower panels) dust continuum emission for different models (first column: 4.0 mm; second: 0.2 mm; third: multi-species dust; last: coagulation) at 5000 or 1000 orbits indicated by the figure labels. Note that the coagulation model only has 1000 orbits due to its computational expense. }\label{fig:image}
\end{figure*}

\subsection{Observational Implications}

For the purpose to compare with (sub-)mm continuum observations,  we utilize \texttt{RADMC-3D} package \citep{Dullemond2012} to produce the 1.3 mm and 7.0 mm dust continuum, and convolve them with  a gaussian beam of $0.03^{\prime\prime}\times0.03^{\prime\prime}$. The $2\ M_{\odot}$ star is assumed to have a black body temperature of 5500 K, and the disk is assumed to be at a distance of $140\ {\rm pc}$.  The details are presented in \citet{Li2019a,Li2019b}.  

The images for different models are shown in Figure~\ref{fig:image}. When the large-scale vortex is sustained, it appears as lopsided horseshoe structures. Here we mainly focus on the long term evolution after the destruction of the gaseous vortex (i.e., $t=5000$ orbits, or 0.3 Myrs). For the $a=4.0\ {\rm mm}$ run,  the dust collected in the vortex is split into several hot spots at both wavelengths. While the vortex is dispersed into a remarkable ring for the $a=0.2\ {\rm mm}$ model \citep{Surville2016,Surville2019}. It is straightforward to understand that the disk is relatively brighter at 1.33 band, and dimmer at 7.0 mm band for the smaller particle size model, as compared with another model.

For the coagulation model, two hot spots are inlaid in a dusty ring, see also the dust distribution in Figure~\ref{fig:coag}. The contrast of the two hot spots becomes more remarkable at 7.0 mm due to the existence of larger particles. Note that we only have 1000 orbits for the coagulation run.
To examine the long term evolution of the structure, we include 5 dust species between $1.0\ \mu{\rm m}$ and $4.0\ {\rm mm}$ with an MRN distribution to mimic, to some extent, multiple dust species.  The large-scale vortex is destroyed $\sim2000$ orbits, which is in between the two single species runs but still longer than that in the coagulation model due to the lack of efficient coagulation/fragmentation. The images at 5000 orbits for two bands are shown in the third column of Figure~\ref{fig:image}.  We can see that the vortex appeared in the early stage will finally become a ring with some remarkable inhomogeneities at 1.33 mm.  Multiple hot spots become prominent at 7.0 mm as the  $a=4.0\ {\rm mm}$ model. It suggests that the features of hot spots inlaid within the ring is mainly related to the coexistence of both large and small particles; the large particles appear observationally as clumpy structures, while small particles can lead to the appearance of the ring. 
This could have gained some observational support for a transition disk LkCa 15, which is shown as some clumpy spots at 7.0 mm band, and appeared as a ringed structure at a shorter wavelength \citep{Andrews2011,Isella2014}.

\section{Conclusions and Discussion}\label{sec:conclusions}

In this work, we perform 2D high-resolution hydrodynamical simulations with LA-COMPASS \citep{Li2005,Li2009,Fu2014b,Li2019b} to study the effect of dust coagulation on the evolution of vortices induced by a massive planet embedded in a low viscosity disk. The dust feedback has also been included to study how the dust coagulation can be effective to maintain or destroy the planet-induced vortices. We have run two single species models with the dust size of $a=4.0\ {\rm mm}$ and $0.2\ {\rm mm}$ to compare with the full coagulation model. A multi-species run is also produced to mimic the long term evolution of the vortex in the coagulation run.

For our coagulation model, due to the
higher gas surface density and steeper pressure gradients within the vortex, dust coagulation/fragmentation and drift to the vortex center are all
quite efficient, producing dust particles ranging from micron to $\sim 1.0\ {\rm cm}$,
as well as overall high $\Sigma_{\rm d}/\Sigma_{\rm g}$ ($\gtrsim1$).   In addition, the dust size distribution is quite spatially non-uniform inside the vortex, 
with the $\Sigma_{\rm d}-$weighted average dust size at the vortex center ($\sim 4.0$ mm)
being a factor of $\sim10$ larger than other vortex regions.  We further find that the gaseous vortex can be destroyed within 1000 orbits of our simulations, which shows a slightly longer lifetime compared to the $a=4.0\ {\rm mm}$ due to the size growth process. The feedback in the coagulation model can become efficient after the total $\Sigma_{\rm d}/\Sigma_{\rm g}\gtrsim1$. It is attributed to the coagulation-assisted drift and then the enhancement of the dust-to-gas ratio of small particles by fragmenting big ones, with the small particles and large ones contributing a comparable fraction to the the total feedback forces. 
Both of them boost the total feedback forces,
which finally destroys the large-scale vortex by the vortex streaming and ``heavy core" instabilities. For the single species run with a large dust size, the effectiveness of dust feedback requires a much higher $\Sigma_{\rm d}/\Sigma_{\rm g}$ by the efficient collecting of particles within the vortex.

We have examined the 1.33 mm and 7.0 mm dust continuum for different models after the vortex is destroyed. For the single species run, 
the dusty structures are manifested as multiple hot spots or a ring depending on the dust size after the gaseous vortex is destroyed. For the coagulation and multi-dust runs, 
several hot spots inlaid in a ring show up in observations, contrary to single species results, as they results from a combination of small and large particles in the ringed structure. Observational support for the coexistence of clumpy and ringed structures has been shown for transition disks (e.g., LkCa 15) at different wavelengths \citep{Andrews2011,Isella2014}.
 
We have not included disk-self gravity based on the initial high minimum Toomre $Q$ ($\sim150$) for disk we have explored. Disk-self gravity could be important for the concentrated dust after the vortex is formed. We have calculated the dust mid-plane density $\rho_{\rm d}$ in the vortex center region for different runs. We find that $\rho_{\rm d}$ for our coagulation run and the small dust size $a=0.2\ {\rm mm}$ run are about two orders of magnitude smaller than the corresponding Roche density $\rho_{\rm R}$, which justifies the neglecting of disk self-gravity. For the larger dust size model, $a=4.0\ {\rm mm}$, $\rho_{\rm d}>\rho_{\rm R}$. We then test this model with the disk self-gravity included. It shows that it can speed up the gaseous vortex evolution, and the similar dusty hot spot structures appear at the later stage.

Note that we have not explored the effect of different coagulation models, planet mass, and disk parameters on the evolution of vortices due to its computational expense. Based on our some preliminary analysis, we expect a power-law disk with a shallow gas profile, which slow down the dust radial drift, could allow the gaseous vortices survival slightly longer. A lower fragmentation velocity and a lower disk mass resulting in a smaller dust size could be helpful to sustain the large-scale vortices to a much longer time. 
This is because the dust feedback effect becomes weaker as the dust size gets smaller. We have tested another single species run with $a=0.02$ mm, and find that the vortex can be sustained to more than 20000 orbits, close to 1 Myr. 
In addition, compared to the 2D model, back reactions are likely less efficient in 3D \citep{Lyra2018}. All of these could extend the lifetime of the gaseous vortex to $\sim$Myrs, and explain the horseshoe structures in (sub)mm observations.  These effect could be explored in details in the future.

\acknowledgments
We thank the referee for useful comments. Y.P.L., H.L., and S.L. gratefully acknowledge the support by LANL/CSES and NASA/ATP.
T.B., J.D., and S.S. acknowledge funding from the European Research Council (ERC)
under the European Union's Horizon 2020 research and innovation
program under grant agreement No. 714769 and support from the Deutsche Forschungsgemeinschaft (DFG, German Research Foundation) through Research Unit
“Transition Disks” (FOR 2634/1, ER 685/8-1) and under Germany's Excellence Strategy – EXC-2094 – 390783311. 
Y.P.L. thanks Daniel Carrera for helpful discussions.
This research used resources provided by the Los Alamos National Laboratory Institutional Computing Program, which is supported by the U.S. Department of Energy National Nuclear Security Administration under Contract No. 89233218CNA000001.

\bibliography{references}{}
\bibliographystyle{aasjournal}



\end{document}